\documentclass[prd,superscriptaddress,nofootinbib,twocolumn,floatfix]{revtex4}
\usepackage{amsfonts}
\usepackage{graphicx,color,amsmath,amssymb}

\def\blfootnote{\xdef\@thefnmark{}\@footnotetext}

\begin{document}

\title{On the non-universality of heavy quark hadronization in elementary high-energy collisions}
\author{Yuxuan Dai}
\affiliation{Department of Applied Physics, Nanjing University of Science and Technology, Nanjing 210094, China}
\author{Shouxing Zhao}
\affiliation{Department of Applied Physics, Nanjing University of Science and Technology, Nanjing 210094, China}
\author{Min He}
\affiliation{Department of Applied Physics, Nanjing University of Science and Technology, Nanjing 210094, China}
\affiliation{Shanghai Research Center for Theoretical Nuclear Physics, NSFC and Fudan University, Shanghai 200438, China}
\date{\today}

\begin{abstract}
It has been traditionally hypothesized that the heavy quark (charm, $c$ and bottom, $b$) fragmentation is universal across different collision systems, based on the notion that hadronization as a soft process should occur at the characteristic non-perturbative QCD scale, $\Lambda_{QCD}$. However, this universality hypothesis has recently been challenged by the observation that the $c$- and $b$-baryon production relative to their meson counterparts in minimum bias proton-proton ($pp$) collisions at the LHC energies is significantly enhanced as compared to the electron-positron ($e^+e^-$) collisions. The conception of non-universality is unambiguously reinforced by the latest measurement of the charged-particle multiplicity dependence of the $b$-baryon-to-meson yield ratio, $\Lambda_b/B$, by the LHCb experiment in $\sqrt{s}=13$\,TeV $pp$ collisions at the LHC, evolving continuously from the saturation value in minimum bias $pp$ collisions toward the small value in $e^+e^-$ collisions as the system size gradually reduces. We address the multiplicity dependence of $b$-baryon production in the canonical statistical hadronization model with input $b$-hadron spectrum augmented with many hitherto unobserved states from quark model predictions. We demonstrate that the decreasing trend of the $\Lambda_b/B$ toward low multiplicities can be quantitatively understood from the canonical suppression on the yield of $\Lambda_b$, as caused by the requirement of strict conservation of baryon number in sufficiently small systems. We have therefore proposed a plausible scenario for understanding the origin of the non-universality of heavy quark fragmentation in elementary collisions.

\end{abstract}


\maketitle

\section{Introduction}
\label{sec_intro}
Owing to their large masses, the production of heavy quarks (charm, $c$, and bottom, $b$) are arguably separated from their hadronization, as implied in the QCD factorization theorem describing the production cross sections of heavy hadrons at collider energies~\cite{Collins:1989gx}. While the heavy quark partonic cross section can be computed as perturbative series in powers of the coupling constant, the hadronization, usually termed fragmentation in elementary collisions, is an intrinsically soft process and thus relies on phenomenological modelling~\cite{Norrbin:2000zc}. Fragmentation fractions of heavy quarks into weakly decaying heavy hadrons (including feeddown contributions from excited states via strong or electromagnetic decays) and ratios between their production yields provide critical probes of the pertinent hadronization mechanisms. Traditionally these fractions (and thus ratios) are thought to be universal across different colliding systems and energies~\cite{Lisovyi:2015uqa}.

However, recent measurements of the charm baryon-to-meson production ratios ($\Lambda_c/D$, $\Sigma_c/D$ and $\Xi_c/D$) by ALICE experiment in minimum bias $pp$ collisions at the LHC energies have revealed a significant enhancement at low transverse momentum ($p_T$) relative to the corresponding values in $e^+e^-$ collisions~\cite{ALICE:2020wfu,ALICE:2021rzj,ALICE:2021bli}, therefore challenging the universality hypothesis. A significantly larger $b$ baryon-to-meson yield ratio ($\Lambda_b/B$) at low $p_T$ compared to $e^+e^-$ collisions has also been observed by the LHCb experiment~\cite{LHCb:2019fns,LHCb:2023wbo} in $pp$ collisions and earlier in $p\bar{p}$ collisions at the Tevatron~\cite{HFLAV:2019otj}. To account for the enhancement of these ratios in hadronic collisions, a statistical hadronization model (SHM) was put forward by assuming {\it relative} chemical equilibrium between the production yields of different heavy hadron species and augmenting the $c$ or $b$-baryon mass spectrum with many hitherto unobserved but theoretically predicted excited states~\cite{He:2019tik,He:2022tod}; the observed enhancement of the $\Lambda_c$ and $\Lambda_b$ production was attributed to the feeddown from these ``missing" states. The measured $\Lambda_c/D$ in $pp$ collisions was also explained by PYTHIA8 calculations that include new color-reconnection ``junctions" fragmenting into baryons~\cite{Christiansen:2015yqa} and by a coalescence model in the presence of a deconfined partonic droplet~\cite{Minissale:2020bif}.

The charged-particle-multiplicity dependent measurements provide a natural means to quantify the evolution of these ratios from $e^+e^-$ to minimum bias $pp$ collisions, and thus characterize how the fragmentation of heavy quarks depends on the density of partons involved. This has been undertaken by ALICE experiment in $pp$ collisions at $\sqrt{s}=13$\,GeV, where $\Lambda_c^+/D^0$ was measured as a function of charged-particle pseudorapidity density $\langle dN_{\rm ch}/d\eta\rangle$ at mid-rapidity and shown to exhibit a significant increase from the lowest to highest multiplicity intervals for $1<p_T<12$\,GeV~\cite{ALICE:2021npz}. The same endeavor has been recently also pursued by the LHCb experiment in the $b$ sector in $\sqrt{s}=13$\,GeV $pp$ collisions at forward rapidity; both $B_s^0/B^0$~\cite{LHCb:2022syj} and $\Lambda_b^0/B^0$~\cite{LHCb:2023wbo} were measured and demonstrated to increase with multiplicity when limited to relatively low $p_T$. In particular, the high-quality data of $p_T$-integrated $\Lambda_b^0/B^0$ converges to the value measured in $e^+e^-$ collisions at the lowest multiplicity and increases by a factor of $\sim$2 toward the saturation value at high multiplicities~\cite{LHCb:2023wbo}, providing an ideal chance of looking into the evolution of heavy quark fragmentation versus the density of the hadronic environment.

To account for the system size dependence of the $\Lambda_c^+/D^0$, it was noted that strict conservation of quantum charges, {\it e.g.,} baryon number, is important when the multiplicity becomes sufficiently low; therefore, one should turn to the version of canonical ensemble of statistical hadronization model~\cite{Chen:2020drg}. Indeed, the {\it strict} baryon number conservation would require the simultaneous production of an antibaryon when creating a baryon, which is very expensive in term of energy, given that the lightest baryon (proton) already has a large mass $\sim 1$\,GeV (compared to the typical hadronization temperature $T_H\sim 160$-170\,MeV). Consequently, the production of the heavy baryons suffers from a canonical suppression, resulting in a decreasing baryon-to-meson ratio toward low multiplicities.

In the present work, we generalize the canonical treatment of the statistical hadornization to the $b$ sector, utilizing the augmented $b$-hadron spectrum that proved indispensable for the statistical interpretation of the enhancement of $\Lambda_b/B$ in minimum bias $pp$ collisions relative to the $e^+e^-$ case~\cite{He:2022tod}. We compute the chemical factors and thermal densities of $b$ hadrons from the canonical partition function and demonstrate the system size dependence of the production of $B_s$ mesons and $\Lambda_b$ and $\Xi_b$ baryons relative to that of $B$ mesons, for both the integrated and $p_T$ differential yields. We show that, with the $b$-hadron input spectrum augmented with many more excited states (in particular excited baryons) beyond the current measured listings, LHCb data of the multiplicity dependence of $\Lambda_b/B$ can be quantitatively understood from the canonical suppression on $\Lambda_b$ production as caused by the {\it strict} conservation of baryon number toward small multiplicities, thereby unveiling a plausible origin of the non-universality of heavy quark hadronization in elementary collisions that is currently under hot debates~\cite{ALICE:2022wpn,Rossi:2023xqa,Altmann:2024icx}.

\begin{table*}[!t]
\begin{center}
\begin{tabular}{lcccccccccc}
\hline\noalign{\smallskip}
$CF$                & $V_C$=5~${\rm fm}^3$     & 10           &  20        &  30       &  50       &  100       &  200    \\
\noalign{\smallskip}\hline\noalign{\smallskip}
$\bar{B}^0$         & 0.0097194                &  0.023927    &	 0.058660  &  0.094845 &  0.16493  &  0.32591   &  0.56988   \\
$B^-$               & 0.0078259                &  0.021863    &  0.056893  &  0.093168 &  0.16331  &  0.32438   &  0.56858     \\
$\bar{B}_s^0$       & 0.0039920                &  0.013624    &	 0.045935  &  0.082725 &  0.15364  &  0.31546   &  0.56101  \\
$\Lambda_b^0$       & 0.0049325                &  0.014844    &  0.047305  &  0.084415 &  0.15574  &  0.31768   &  0.56300        \\
$\Xi_b^{0-}$        & 0.0021863                &  0.0089128   &	 0.037336  &  0.073498 &  0.14477  &  0.30720   &  0.55402     \\
$\Omega_b^-$        & 0.0004649                &  0.0030092   &	 0.019475  &  0.047296 &  0.11221  &  0.27231   &  0.52265      \\

\noalign{\smallskip}\hline\noalign{\smallskip}
$\bar{B}_s^0/\bar{B}^0$       & 0.41072        & 0.56939      &	 0.78307   &  0.87221  &  0.93155  &  0.96793   &  0.98443	       \\
$\Lambda_b^0/\bar{B}^0$       & 0.50749        & 0.62039      &  0.80643   &  0.89003  &  0.94427  &  0.97474   &  0.98793  \\
$\Xi_b^{0-}/\bar{B}^0$        & 0.22494        & 0.37250      &  0.63648   &  0.77493  &  0.87776  &  0.94259   &  0.97217  \\

\noalign{\smallskip}\hline
\end{tabular}
\end{center}
\caption{Chemical factors (CF) of the direct ground-state $b$-hadrons at $T_H=170$\,MeV, $\gamma_s=0.6$, $\gamma_c=15$ and $\gamma_b=10^9$ for varying correlation volumes, with $b$-hadron input spectrum taken from the RQM scenario. The ratios of the CF's of $B_s^0$, $\Lambda_b^0$ and $\Xi_b^{0-}$ to $\bar{B}^0$ are also summarized in the last three rows.}
\label{tab_CFs-vs-volume}
\end{table*}

\section{Canonical partition function and input bottom-hadron spectrum}
\label{sec_canonical-SHM}

When applied to large systems where the fluctuations of quantum charges are relatively small, SHM is formulated in a grand-canonical ensemble (GCE) of an ideal hadron resonance gas~\cite{Andronic:2017pug}. The Abelian quantum charges, such as electric charge $Q$, baryon number $N$, strangeness $S$, charm number $C$ and bottom number $B$ are thus conserved {\it on average} and regulated by the corresponding chemical potentials $\vec\mu=(\mu_Q,\mu_N,\mu_S,\mu_C,\mu_B)$. The primary mean yield for the $j$-th hadron produced from GCE-SHM is given by
\begin{align}\label{GCE-SHM}
\langle N_j\rangle^{GCE}=\gamma_s^{N_{sj}}\gamma_c^{N_{cj}}\gamma_b^{N_{bj}}z_je^{\vec\mu\cdot\vec q_j/T_H},
\end{align}
where $\vec q_j=(Q_j,N_j,S_j,C_j,B_j)$ denote the quantum charges for the hadron and $z_j$ is the one-particle partition function
\begin{align}\label{1-partition-function}
z_j=(2J_j+1)\frac{VT_H}{2\pi^2}m_j^2K_2(\frac{m_j}{T_H}),
\end{align}
which specifies the chemical equilibrium multiplicity of the $j$-th hadron of mass $m_j$ and spin $J_j$ in a fireball of volume $V$ under the Boltzmann approximation at hadronization temperature $T_H$, with $K_2$ being the modified Bessel function of the second order. In Eq.~(\ref{GCE-SHM}), $\gamma_s$, $\gamma_c$ and $\gamma_b$ denote fugacities that account for the deviation from chemical equilibrium for hadrons containing $N_{sj}$, $N_{cj}$ and $N_{bj}$ strange, charm and bottom quarks or antiquarks, respectively.

In contrast, for sufficiently small systems where relative fluctuations of quantum charges become significant, one has to turn to the canonical ensemble (CE) SHM, in order to {\it strictly} conserve the quantum charges~\cite{Rafelski:1980gk,Hagedorn:1984uy,Vovchenko:2019kes,Cleymans:2020fsc,Chen:2020drg}. For a system having conserved quantum charges $\vec Q=(Q,N,S,C,B)$ with associated phase angles $\vec\phi=(\phi_Q,\phi_N,\phi_S,\phi_C,\phi_B$), the CE-SHM partition function reads
\begin{align}\label{CE-SHM}
Z(\vec Q)=\int_0^{2\pi}\frac{d^5\phi }{(2\pi)^5}e^{i\vec Q\cdot\vec\phi}{\rm exp}[\sum_j\gamma_s^{N_{sj}}\gamma_c^{N_{cj}}\gamma_b^{N_{bj}}e^{-i\vec q_j\cdot\vec\phi}z_j],
\end{align}
where the volume in $z_j$ should now be understood as the correlation volume $V_C$ that characterizes the range of the {\it strict} conservation of quantum charges. In Eq.~(\ref{CE-SHM}), the summation $\sum_j$ should be taken over {\it all} hadrons up to the most massive $b$-mesons and -baryons, for the present aim of investigating the statistical hadronization of $b$ quarks. More specifically, all light hadrons in the particle data group (PDG) listings~\cite{ParticleDataGroup:2022pth} are included. For $c$ hadrons, in particular $c$ baryons, excited states from relativistic quark model (RQM) predictions~\cite{Ebert:2009ua,Ebert:2011kk} beyond the current PDG listings are included, which have proved essential for interpreting the observed enhancement of $\Lambda_c/D$ in minimum bias $pp$ collisions in the GCE-SHM~\cite{He:2019tik}. All charmonium states listed in PDG are also included. For $b$-hadron spectrum serving as the direct input for the present study, we compare two sets of $b$ hadrons in the same spirit as in the GCE-SHM study for minimum bias collisions~\cite{He:2022tod}: (a) PDG-only states~\cite{ParticleDataGroup:2022pth} and (b) RQM states~\cite{Ebert:2009ua,Ebert:2011kk} that additionally include 18 $B$'s, 16 $B_s$'s, 27 $\Lambda_b$'s, 45 $\Sigma_b$'s, 71 $\Xi_b$'s and 41 $\Omega_b$'s, up to meson (baryon) masses of 6.5 (7.0) GeV.

The primary mean yield of the $j$-th hadron produced from the CE-SHM then reads~\cite{Chen:2020drg,Becattini:2009sc}
\begin{align}\label{CE-SHM-yields}
\langle N_j\rangle^{CE}=\gamma_s^{N_{sj}}\gamma_c^{N_{cj}}\gamma_b^{N_{bj}}z_j\frac{Z(\vec Q-\vec q_j)}{Z(\vec Q)}.
\end{align}
The {\it chemical factor} $Z(\vec Q-\vec q_j)/Z(\vec Q)$ in favor of the chemical potential term in Eq.~(\ref{GCE-SHM}) arises from the requirement of {\it exact} conservation of quantum charges. As will be shown in the following, for a completely neutral system ({\it i.e.}, $\vec Q=0$ and thus $\vec\mu=0$), this factor for charged hadrons with $\vec q_j\neq0$ is always less than unity but only tends to unity at asymptotically large volumes, therefore characterizing the {\it canonical suppression} of the production of charged hadrons at small system size.

\section{System size dependence of $b$-hadron yields}
\label{sec_computations}
The measurements of charged particle multiplicity dependence of $B_s/B$~\cite{LHCb:2022syj} and $\Lambda_b/B$~\cite{LHCb:2023wbo} by the LHCb experiment in $\sqrt{s}=13$\,TeV $pp$ collisions were performed at forward rapidity ($2<y<4.5$), where particle and anti-particle production is not totally symmetric. However, the measured asymmetry between $\Lambda_b^0$ and $\bar{\Lambda}_b^0$ is at the level of $\sim$1-2$\%$ in $\sqrt{s}=7$\,TeV $pp$ collisions at $2<y<4.5$, and becomes even smaller in $\sqrt{s}=8$\,TeV collisions~\cite{LHCb:2021xyh}. Therefore, as a good approximation, we focus on a completely neutral system with vanishing quantum charges ($\vec Q=(Q,N,S,C,B)=(0,0,0,0,0)$) in the following CE-SHM studies, with RQM being the default scenario for $b$-hadron spectrum input.

\subsection{Chemical factors of direct $b$ hadrons}
\label{ssec_CFs}

\begin{table*}[!t]
\begin{center}
\begin{tabular}{lcccccccccc}
\hline\noalign{\smallskip}
$n_{\alpha}(\cdot10^{-5}~{\rm fm^{-3}})$   & $V_C$=5~${\rm fm}^3$ & 10   &  20        &  30       &  50       &  100       &  200      & GCE\\
\noalign{\smallskip}\hline\noalign{\smallskip}
$\bar{B}^0$         & 1.1220                               &  2.7920      &	 6.9508    &  11.313   &  19.759   &  39.148    &  68.534  & 120.41  \\
$B^-$               & 0.96934                              &  2.6261      &  6.8105    &  11.181   &  19.635   &  39.038    &  68.452  & 120.45 \\
$\bar{B}_s^0$       & 0.14641                              &  0.47267     &	 1.5299    &  2.7242   &  5.0273   &  10.285    &  18.263  & 32.513  \\
$\Lambda_b^0$       & 0.29886                              &  0.90201     &  2.8845    &  5.1551   &  9.5210   &  19.435    &  34.453  & 61.702  \\
$\Xi_b^{0-}$        & 0.043883                             &  0.17479     &	 0.72393   &  1.4247   &  2.8132   &  5.9882    &  10.818  & 19.548  \\
$\Omega_b^-$        & 0.00028060                           &  0.0018164   &	 0.011755  &  0.028549 &  0.067730 &  0.16437   &  0.31548 & 0.63204 \\

\noalign{\smallskip}\hline\noalign{\smallskip}
$\bar{B}_s^0/\bar{B}^0$       & 0.13049                    &  0.16929     &  0.22010   &  0.24080  &  0.25443  &  0.26273   &  0.26648 & 0.27002 \\
$\Lambda_b^0/\bar{B}^0$       & 0.26635                    &  0.32307     &  0.41499   &  0.45568  &  0.48186  &  0.49644   &  0.50271 & 0.51243 \\
$\Xi_b^{0-}/\bar{B}^0$        & 0.039110                   &  0.062602    &  0.10415   &  0.12594  &  0.14238  &  0.15296   &  0.15785 & 0.16235 \\

\noalign{\smallskip}\hline
\end{tabular}
\end{center}
\caption{Total ({\it i.e.}, with feeddowns) thermal densities of ground-state $b$-hadrons at $T_H=170$\,MeV, $\gamma_s=0.6$, $\gamma_c=15$ and $\gamma_b=10^9$ for varying correlation volumes, with $b$-hadron input spectrum taken from the RQM scenario. The ratios of the thermal densities of $B_s^0$, $\Lambda_b^0$ and $\Xi_b^{0-}$ to $\bar{B}^0$ are also summarized in the last three rows. The last column denotes the values in the grand canonical limit.}
\label{tab_densities-vs-volume}
\end{table*}

To evaluate the canonical partition function, hadrons to be summed over in the exponential of Eq.~(\ref{CE-SHM}) are divided into three categories~\cite{Chen:2020drg}: the completely neutral mesons with $\vec q_j=(Q_j,N_j,S_j,C_j,B_j)=(0,0,0,0,0)$, {\it e.g.}, $\pi^0$, $\rho^0$, $\phi$ and $J/\psi$; the charged mesons, including positively charged mesons with at least one of $Q_j=+1,S_j=+1,C_j=+1,B_j=+1$ but $N_j=0$, {\it e.g.}, $\pi^+$, $K^+$, $D^+$, $D_s^+$, $B^+$, $B_s^0$ and their antimesons; the baryons with $N_j=+1$ and antibaryons with $N_j=-1$. The hadronization temperature is taken to be $T_H=170$\,MeV~\cite{He:2022tod}; as we've checked for the GCE-SHM study~\cite{He:2022tod}, using a lower $T_H\sim160$\,MeV leads to similar results for the $b$-hadron fragmentation fractions. To proceed, the strangeness under-saturation parameter (fugacity) is fixed to be the typical value $\gamma_s=0.6$ in elementary collisions~\cite{He:2019tik,He:2022tod}, which has been shown to be insensitive to the system size for charged particle multiplicity $dN_{ch}/d\eta\leq 50$~\cite{Vovchenko:2019kes}. For the $c$ and $b$ fugacity, $\gamma_c$ and $\gamma_b$ respectively, that measure their initial hard production in excess of the chemical equilibrium limit, we take the typical value of $\gamma_c\sim15$~\cite{Chen:2020drg} and $\gamma_b\sim10^9$ determined in the canonical SHM study of semi-central heavy-ion collisions at the LHC energy~\cite{Andronic:2021erx,Andronic:2022ucg}. We've checked that our final result of the $b$-baryon relative to -meson production is rather robust against a significant variation of $\gamma_b$ within four orders of magnitude around the default value used here, since anyway the rather massive $b$ hadrons only account for a tiny contribution to the canonical partition function (through summation in the exponential of Eq.~(\ref{CE-SHM})).

In Table~\ref{tab_CFs-vs-volume}, chemical factors (CF) of the direct ground state $b$ hadrons are displayed for varying correlation volumes ($V_C$). At small $V_C$, $\bar{B}_s^0$ and $\Lambda_b^0$ exhibit significantly smaller CF's than $\bar{B}^0$ and $B^-$ (we focus on mesons and baryons both containing a $b$ quark, instead of the antiquark $\bar{b}$), owing to the canonical suppression brought about by the {\it exact} strangeness and baryon number conservation, respectively. Despite all being $b$-baryon, the CF's of $\Xi_b^{0-}$ and $\Omega_b^-$ are further progressively smaller compared to that of $\Lambda_b^0$, as a result of their increasing strangeness content leading to stronger canonical strangeness suppression. As the correlation volume increases, the canonical suppression effects induced by the
{\it exact} strangeness and baryon-number conservation attenuate, and hence at asymptotically large volume, the CF's of all $b$ hadrons end up with almost the same residual value $\sim0.56$ that arises solely from their {\it common} canonical bottom-number suppression. This is also seen from the gradual increase of the {\it relative} canonical suppression ({\it i.e.}, the ratios of CF's listed in the last three rows of Table~\ref{tab_CFs-vs-volume}) $\bar{B}_s^0/\bar{B}^0$, $\Lambda_b^0/\bar{B}^0$ and $\Xi_b^{0-}/\bar{B}^0$ toward unity from small to large volumes.

\subsection{Total thermal densities of ground-state $b$ hadrons}
\label{ssec_densities}

\begin{figure} [!t]
\includegraphics[width=1.05\columnwidth]{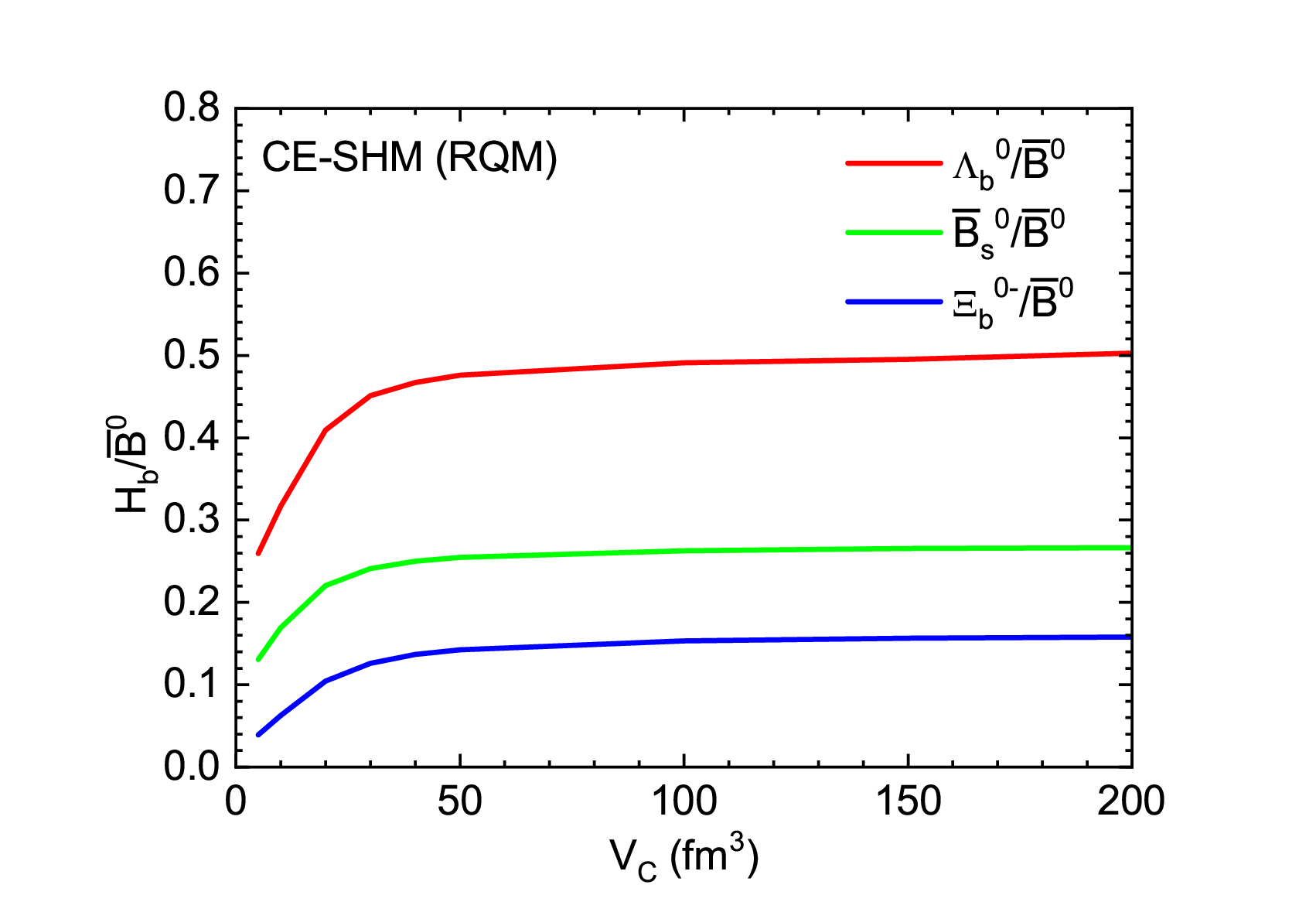}
\vspace{-0.3cm}
\caption{The correlation volume dependence of the ratios of the thermal densities of $\bar{B}_s^0$, $\Lambda_b^0$, and $\Xi_b^{0-}$ to that of $\bar{B}^0$ from CE-SHM with $b$-hadron input spectrum taken from the RQM scenario.}
\label{Lb-Bs-Xib-to-B-vs-VC}
\end{figure}

With the primary mean yields of $b$ hadrons computed from Eq.~(\ref{CE-SHM-yields}), the total thermal densities of ground state $b$ hadrons are obtained from the sum of the direct one and the feeddown contributions from excited states
\begin{equation}\label{total-density}
n_{\alpha}=\frac{\langle N_{\alpha}\rangle^{CE}}{V_C} + \sum_j\frac{\langle N_j\rangle^{CE}}{V_C}\cdot{\rm BR}(j\rightarrow\alpha),
\end{equation}
where the branching ratios (BR) for excited $b$ hadrons decaying to the ground states were given from a $^3P_0$ quark model estimate~\cite{He:2022tod}. These densities are shown in Table~\ref{tab_densities-vs-volume} for varying correlations volumes, together with the GCE-SHM results. An immediate observation is that as the volume increases, the thermal density for each $b$-hadron also grows. While these thermal densities at the largest volume ($V_C=200~{\rm fm}^3$) computed here are still far from the GCE-SHM values, the corresponding ratios between them are already almost the same as the grand-canonical limiting values, simply because the effects of canonical strangeness and baryon-number suppression already become vanishing and the residual canonical bottom-number suppression is {\it common} for all $b$ hadrons. As the most important finding of the present work, the thermal density ratios $\bar{B}_s^0/\bar{B}^0$, $\Lambda_b^0/\bar{B}^0$ and $\Xi_b^{0-}/\bar{B}^0$, as summarized in the last three rows of Table~\ref{tab_densities-vs-volume} and also plotted in Fig.~\ref{Lb-Bs-Xib-to-B-vs-VC}, demonstrate a marked system-size dependence, amounting to a factor of 2-4 reduction from the saturation value in the grand-canonical limit to the smallest volume and serving as a direct signal of the canonical suppression effects due to {\it strict} conservation of strangeness and baryon number.

\subsection{Confronting LHCb data}
\label{ssec_confronting-data}

\begin{figure} [!t]
\includegraphics[width=1.05\columnwidth]{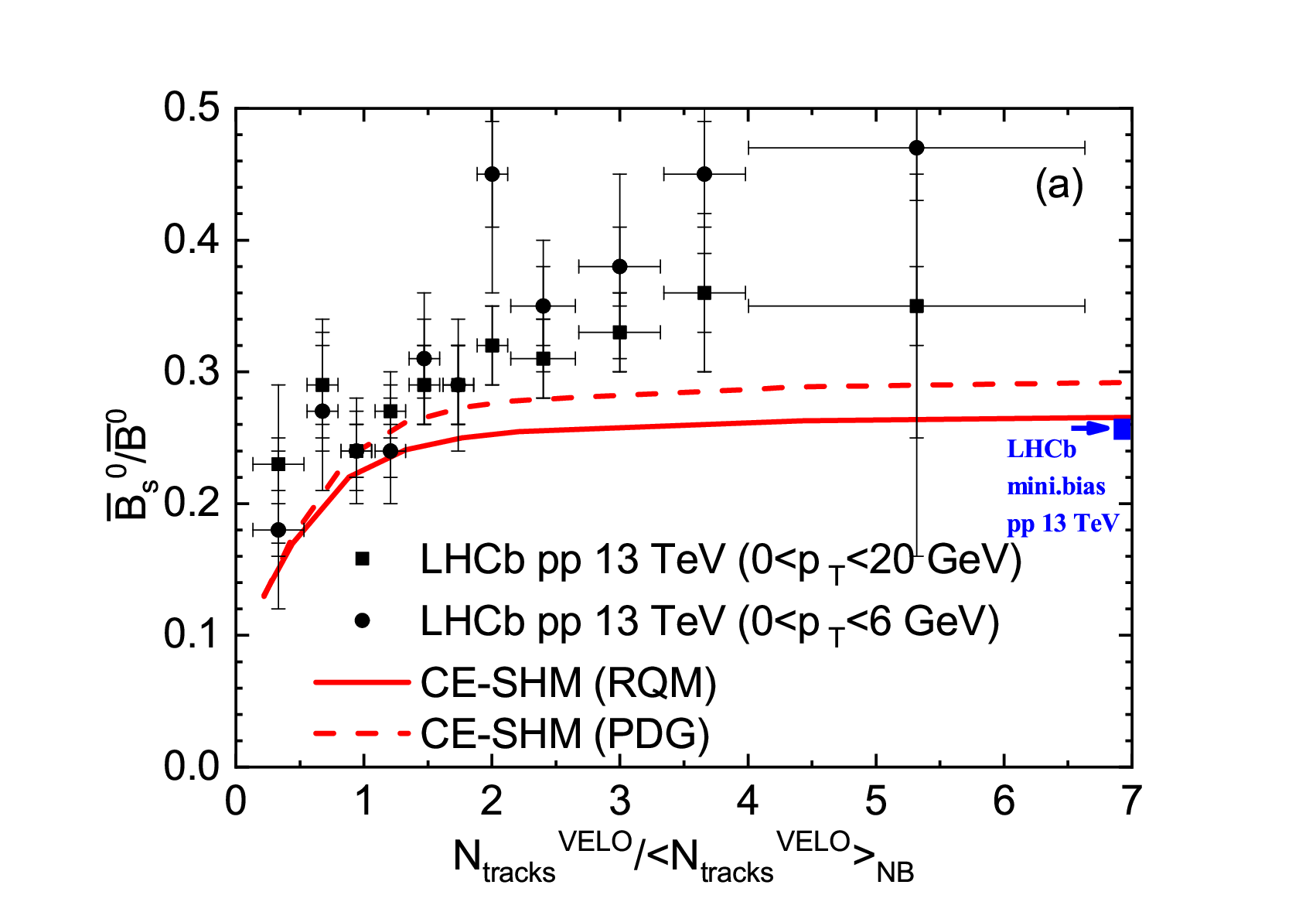}
\vspace{-0.3cm}
\includegraphics[width=1.05\columnwidth]{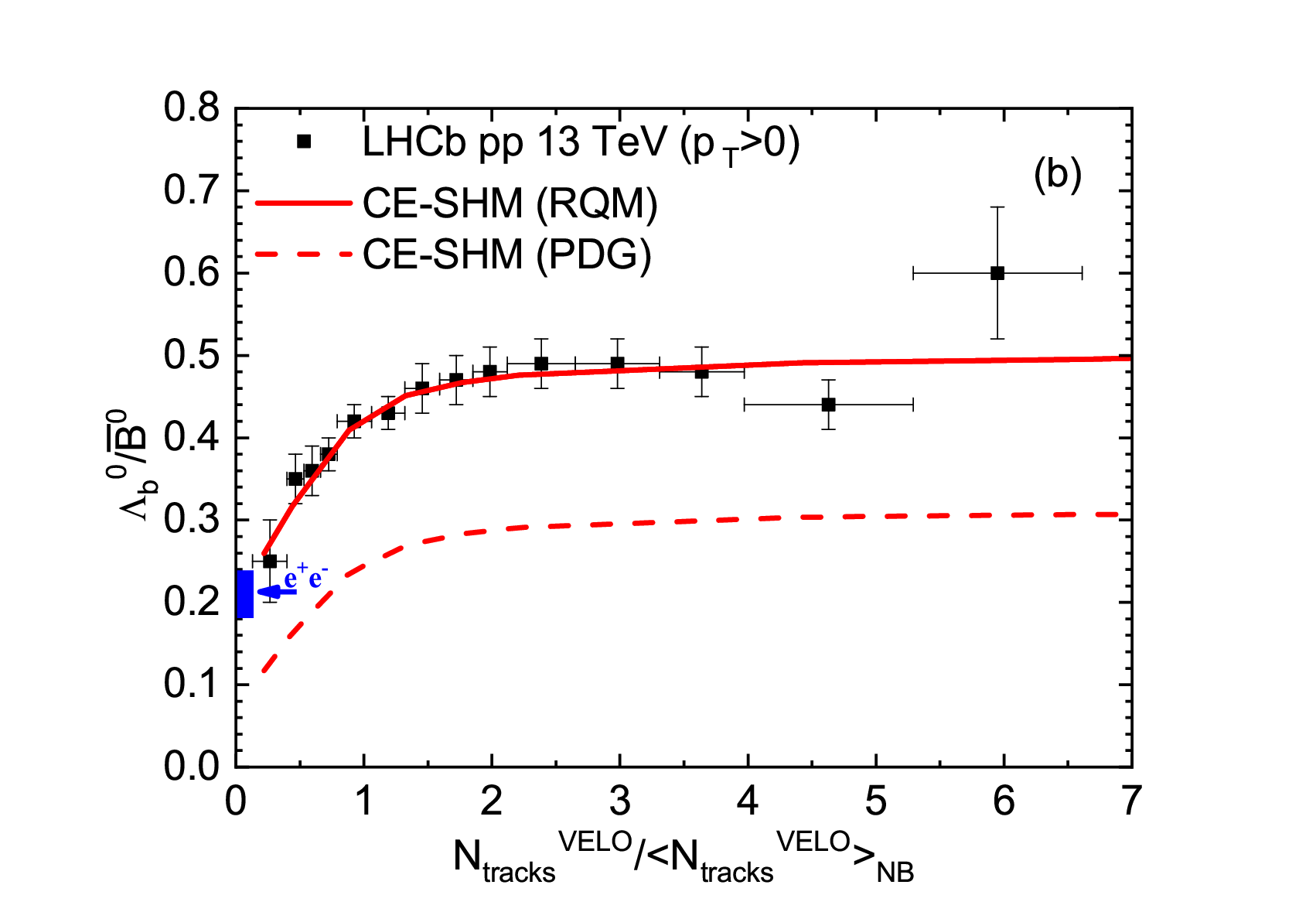}
\vspace{-0.3cm}
\caption{The system size dependence of the yield ratios of (a) $\bar{B}_s^0/\bar{B}^0$ and (b) $\Lambda_b^0/\bar{B}^0$ as predicted from CE-SHM with $b$-hadron input spectrum taken from the RQM (solid lines) versus PDG (dashed lines) scenario, in comparison with LHCb data~\cite{LHCb:2022syj,LHCb:2023wbo}. The filled box at the right vertical axis of the upper panel indicate a separate measurement of $\bar{B}_s^0/\bar{B}^0=0.2539\pm 0.0079$ in minimum bias $pp$ collisions at the same $\sqrt{s}=13$\,TeV by the LHCb experiment~\cite{LHCb:2021qbv}, and the filled box at the left vertical axis of the lower panel indicates the $\Lambda_b^0/\bar{B}^0$ measured in $e^+e^-$ collisions~\cite{HFLAV:2019otj,LHCb:2023wbo}.}
\label{Bs-Lb-to-B-vs-LHCb-data}
\end{figure}

In order to compare the system size dependence of the $\bar{B}_s^0/\bar{B}^0$, $\Lambda_b^0/\bar{B}^0$ as uncovered from the present GC-SHM study with the LHCb measurements~\cite{LHCb:2022syj,LHCb:2023wbo}, where the system size was quantified by the total number of charged tracks, we follow a previous canonical thermal study of the light hadrons that suggests a linear dependence between the correlation volume and the measured charged particle multiplicity~\cite{Vovchenko:2019kes}. We find that assuming a mean correlation volume $\langle V_C\rangle=22.6~{\rm fm}^3$ corresponding to the mean number of tracks $\langle N_{\rm tracks}^{\rm VELO}\rangle_{\rm NB}$, one could establish a simple relation $V_C/\langle V_C\rangle=N_{\rm tracks}^{\rm VELO}/\langle N_{\rm tracks}^{\rm VELO}\rangle_{\rm NB}$ that translates the predicted volume dependence of $\Lambda_b^0/\bar{B}^0$ into a best fit of the LHCb data, as shown in Fig.~\ref{Bs-Lb-to-B-vs-LHCb-data}(b) together with $\bar{B}_s^0/\bar{B}^0$ (Fig.~\ref{Bs-Lb-to-B-vs-LHCb-data}(a)), where we also compare the RQM versus PDG scenario as the $b$-hadron spectrum input. While the current LHCb data of the system size dependence of $\bar{B}_s^0/\bar{B}^0$ do not allow for a discrimination between the PDG and RQM scenario (however the minimum bias datum indicated by the filled box at the right vertical axis in panel (a) prefers the RQM scenario), data of $\Lambda_b^0/\bar{B}^0$ show an unambiguous preference for the RQM scenario. In particular, the decreasing behavior of the $\Lambda_b^0/\bar{B}^0$ data from the saturation value in minimum bias collisions toward the $e^+e^-$ value at the smallest system size is quantitatively described by our CE-SHM prediction within the RQM scenario. This provides another strong evidence for the existence of many not-yet-observed excited states of $b$ hadrons (particularly baryons)~\cite{He:2022tod}, and suggests a plausible mechanism for the non-universality of heavy quark fragmentation in elementary collisions, namely the canonical suppression of baryon production at sufficiently small system size as caused by the requirement of {\it exact} baryon-number conservation.

\section{Fragmentation simulation: $p_T$ differential ratios}
\label{sec_fragmentation}

\begin{figure} [!t]
\includegraphics[width=1.05\columnwidth]{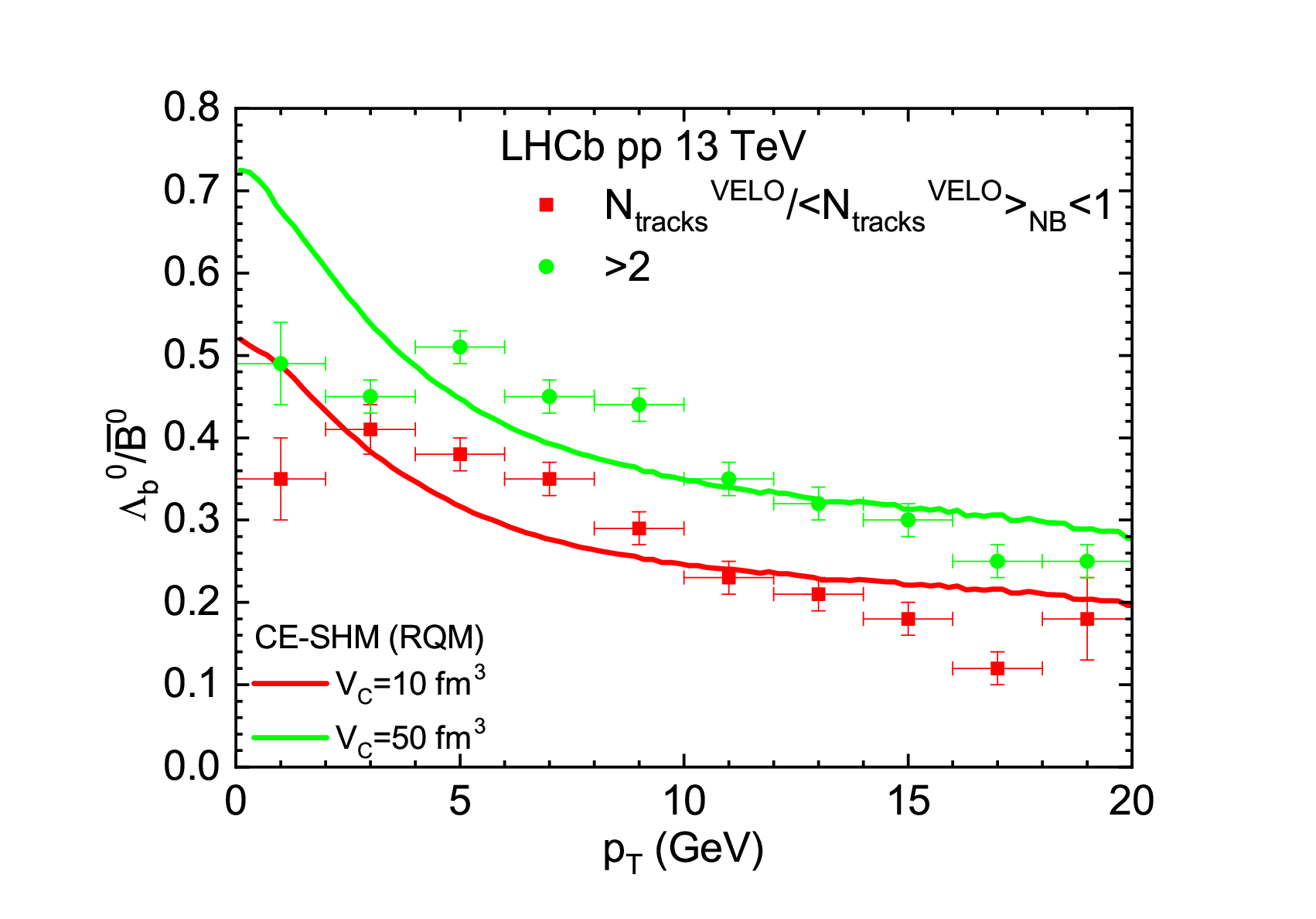}
\vspace{-0.3cm}
\caption{The $p_T$ differential $\Lambda_b^0/\bar{B}^0$ in two multiplicity intervals in comparison with LHCb data~\cite{LHCb:2023wbo}.}
\label{pT-differntial-Lb-to-B}
\end{figure}

To compute the $p_T$ differential spectrum and ratios of ground-state $b$ hadrons, we sample the perturbative $b$ quark $p_t$ spectrum computed from FONLL~\cite{Frixione:2007nw,Cacciari:2012ny} in $\sqrt{s}=13$\,TeV collisions and perform the fragmentation simulation with the fragmentation function
\begin{equation}\label{bFF}
D_{b\rightarrow H_b}(z)\propto z^{\alpha}(1-z) \ ,
\end{equation}
where $z=p_T/p_t$ is the fraction of the $b$-hadron's ($H_b$) transverse momentum, $p_T$, over the $b$-quark one, $p_t$. The fragmentation weight
of $H_b$ is assumed to be proportional to its primary thermal density, $\langle N_{H_b}\rangle^{CE}/V_C$ (Eq.~(\ref{CE-SHM-yields})), in the spirit of {\it relative} chemical equilibrium. Each $H_b$ produced from fragmentation is then decayed into the ground-state particles with a constant matrix element ({\it i.e.}, decay kinematics solely determined by phase space) and ${\rm BR}$'s used in Eq.~(\ref{total-density})~\cite{He:2022tod}. The parameter $\alpha$ in Eq.~(\ref{bFF}) is tuned to fit the slope of the $p_T$ spectrum of ground-state $b$ hadrons. For simplicity, we take $\alpha=45,25,8$ for all $B$ mesons, all $B_s$ mesons and all $b$-baryons, respectively, following Ref.\cite{He:2022tod} for minimum bias collisions. The $p_T$ spectra of $\bar{B}^0$ and $\Lambda_b^0$ thus obtained are divided each other to get the $p_T$ differential ratio of $\Lambda_b^0/\bar{B}^0$, as shown in Fig.~\ref{pT-differntial-Lb-to-B} for two correlation volumes in comparison with LHCb data for the low and high multiplicity intervals. The separation of $\Lambda_b^0/\bar{B}^0$ between these two intervals and its trend of $p_T$ dependence can be qualitatively described.

\section{Summary}
\label{sec_sum}

We have addressed the system size dependence of the production of $b$ baryons relative to that of mesons in high-energy $pp$ collisions. By introducing many hitherto unobserved but theoretically predicted $b$ hadrons (in particular excited baryons) into the canonical ensemble SHM, we've demonstrated that the charged-particle-multiplicity dependence of $\Lambda_b/B$ measured by the LHCb experiment can be quantitatively described, attributed to the canonical suppression on $\Lambda_b$ production toward low multiplicities as a result of {\it exact} conservation of baryon number as required by the canonical treatment of SHM. The $\Lambda_b/B$ as explored bridges continuously the gap between the saturation value in minimum bias $pp$ collisions and the small value in $e^+e^-$ collisions, reinforcing the conception that the heavy quark hadronization depends on the hadronic environment involved in the collision systems. We have therefore proposed a possible origin of the non-universality of heavy quark hadronization in terms of canonical baryon suppression at sufficiently small system size in the presence of many ``missing" baryons awaiting discovery.

\acknowledgments M. H. thanks Profs. Y.-G. Ma, G.-L. Ma and J.-H. Chen for the hospitality during his stay as a visiting scholar at Shanghai Research Center for Theoretical Nuclear Physics of Fudan University where the work was finalized. This work was supported by the National Natural Science Foundation of China (NSFC) under Grants No.12075122 (M. H.) and No.12147101 (via Shanghai Research Center for Theoretical Nuclear Physics).

\end{document}